\documentclass[aps,nofootinbib,a4paper,floats,eqsecnum,12pt,tightenlines]{revtex4}

\usepackage[dvips]{graphicx}
\begin{document}
    
\newcommand{\beq}{\begin{equation}} 
\newcommand{\eeq}{\end{equation}} 
\newcommand{\beqa}{\begin{eqnarray}} 
\newcommand{\eeqa}{\end{eqnarray}} 
\newcommand{\da}{^\dagger} 
\newcommand{\wh}{\widehat}

\newcommand{\Om}{\Omega}
\newcommand{\tr}{{\rm tr}}
\newcommand{\intf}{\int_{-\infty}^\infty}
\newcommand{\into}{\int_0^\infty}
\newcommand{\I}{{\mathcal I}}
\newcommand{\G}{{\mathcal G}}
\newcommand{\A}{{\mathcal A}}
\newcommand{\F}{{\mathcal F}}

\renewcommand{\=}{\!\!=\!\!}

\def\simleq{\; \raise0.3ex\hbox{$<$\kern-0.75em
      \raise-1.1ex\hbox{$\sim$}}\; }
\def\simgeq{\; \raise0.3ex\hbox{$>$\kern-0.75em
      \raise-1.1ex\hbox{$\sim$}}\; }

\def\la{{\langle}} 
\def\ra{{\rangle}} \def\vep{{\varepsilon}} \def\y{\'\i}
\def\half{{1\over 2}}
\def\an{|\Phi _N \rangle }
\def\bn{|\Phi ' _{N'}\rangle }
\def\na{ \langle \Phi _N|}
\def\nb{ \langle \Phi'_{N'} |}

\def\ov{\over}
\def\non{\nonumber }
\def\beq{\begin{equation} }
\def\eeq{\end{equation} }
\def\beqa{\begin{eqnarray}}
\def\eeqa{\end{eqnarray}}
\def\del{\partial }
\def\D{\Delta}
\def\a{\alpha }
\def\am{\alpha^\mu }
\def\Xm{X^\mu}
\def\Xn{X^\nu}
\def\d{\textrm{d}}
\def\b{\beta}
\def\t{\tau}
\def\e{\epsilon}
\def\g{\gamma}
\def\med{\frac{1}{2}}

\def\npb#1#2#3{{Nucl. Phys.} {\bf B#1} (#2) #3}
\def\plb#1#2#3{{ Phys. Lett.} {\bf B#1} (#2) #3}
\def\prd#1#2#3{{Phys. Rev.} {\bf D#1} (#2) #3}
\def\prl#1#2#3{{Phys. Rev. Lett.} {\bf #1} (#2) #3}
\def\mpla#1#2#3{{Mod. Phys. Lett.} {\bf A#1} (#2) #3}
\def\ijmpa#1#2#3{{Int. J. Mod. Phys.} {\bf A#1} (#2) #3}
\def\jmp#1#2#3{{J. Math. Phys.} {\bf #1} (#2) #3}
\def\cmp#1#2#3{{ Commun. Math. Phys.} {\bf #1} (#2) #3}
\def\pr#1#2#3{{Phys. Rev. } {\bf #1} (#2) #3}
\def\umn#1#2#3{{ Usp. Matem. Nauk} {\bf #1} (#2) #3}
\def\bb#1{{\tt hep-th/#1}}
\def\grqc#1{{\tt gr-qc/#1}}
\def\heph#1{{\tt hep-ph/#1}} 
\def\mathph#1{{\tt math-ph/#1}}
\def\ptp#1#2#3{{\it Prog. Theor. Phys.} {\bf #1} (#2) #3}
\def\rmp#1#2#3{{\it Rev. Mod. Phys.} {\bf #1} (#2) #3}
\def\jetplet#1#2#3{{\it JETP Letters} {\bf #1} (#2) #3}
\def\jetp#1#2#3{{\it Sov. Phys. JEPT} {\bf #1} (#2) #3}
\def\pha#1#2#3{{\it Physica } {\bf A#1} (#2) #3 }
\def\app#1#2#3{{\it Astropart. Phys.} {\bf #1} (#2) #3 }
\def\cqg#1#2#3{{\it Class. Quantum Grav.} {\bf #1} (#2) #3 }
\def\apj#1#2#3{{\it Astrophys. J.} {\bf #1} (#2) #3 }
\def\ma#1#2#3{{\it Math. Ann.} {\bf #1} (#2) #3 }
\def\npps#1#2#3{{\it Nucl. Phys. Proc. Suppl.} {\bf #1} (#2) #3 }
\def\atmp#1#2#3{{\it Adv. Theor. Math. Phys.} {\bf #1} (#2) #3 }
\def\jhep#1#2#3{{ J. High Energy Phys.} {\bf #1} (#2) #3}
\def\cm#1{{\tt cond-mat/#1}}
\def\aph#1{{\tt astro-ph/#1}}
\def\nc#1#2#3{{\it Nuovo Cimento } {\bf #1} (#2) #3 }
\def\nat#1#2#3{{\it Nature} {\bf #1} (#2) #3 }
\def\ijtp#1#2#3{{\it Int. J. Theor. Phys.} {\bf #1} (#2) #3 }
\def\prep#1#2#3{{\it Phys. Rep.} {\bf #1} (#2) #3 }
\def\aj#1#2#3{{\it Astron. J.} {\bf #1} (#2) #3 }

\preprint{EHU-FT/03-05}

\title{String Form Factors}


\author{J. L. Ma\~nes}
\affiliation{Departamento de F\y sica  de la Materia Condensada\\
 Universidad del Pa\'{\i}s Vasco,
 Apdo. 644, E-48080 Bilbao, Spain  \\
 {\bf wmpmapaj@lg.ehu.es}}


\date{\today}

\begin{abstract}
We compute the cross section for scattering of light string probes by 
randomly  excited closed strings. For high energy probes, 
the cross section factorizes and  can be used to define 
 effective form factors for the excited targets. These form factors are 
well defined without the need for infinite subtractions and contain 
information about the shape and size of typical  strings. For 
highly excited  strings the elastic form factor can be written 
in terms of   the 
`plasma dispersion function', which describes charge screening in  high temperature 
plasmas. 
\end{abstract}

\pacs{????}

\maketitle

\section{Introduction}

Classically, a string is a one dimensional object and its shape is 
simply given by the set of functions $X^\mu(\sigma,\tau)$. However, when we try to define the 
size and shape of a quantum string we run into trouble, since the 
coordinates
$X^\mu$ become quantum fields on the worldsheet and, as such, undergo 
infinite zero-point fluctuations~\cite{had}. For instance, if we try to compute the 
mean square radius of a ground state string (i.e. a tachyon,  
photon or  graviton, depending on the concrete string theory) we find
\beq
\la X^2\ra\sim\alpha'\sum_{1}^\infty{1\over n}=\alpha' \log 
\infty\, .
\eeq

This, however, is not the end of the story, as was recognized long 
ago~\cite{size}. The reason we get an infinite result for the mean square 
radius is that we are summing over the infinite modes of the string. 
But any real attempt to measure the size of the string will be 
limited by the time resolution $\epsilon$ of the experiment, and all 
modes with frequency $\omega>1/\epsilon$ will be averaged out and 
effectively cut off, so that $\la X^2\ra\sim\a'\log (1/\epsilon)$.

 A very 
natural way to measure the shape of a string is by a scattering 
experiment. Consider for instance the Veneziano amplitude $\A(s,t)$ describing 
the scattering of two open string tachyons. In the Regge limit of fixed momentum 
transfer $t=-q^2$ and high  energy $E$ (with $s\sim 2 E^2$) the amplitude 
can be written as~\cite{gsw,pol}
\beq\label{regge}
\A\sim s^{1-\a' q^2} \Gamma( \a'q^2-1)\,.
\eeq
At low momentum transfer the amplitude is dominated by the photon pole 
in the gamma function and can be interpreted as the product of the photon propagator times 
the form factors $\F(q^2)$ of the scattered strings~\cite{comp}:
\beq
\A\sim {E^2\over q^2} (e^{-\a' q^2 \log E})^2={E^2\over 
q^2}\F^2(q^2)\,.
\eeq
  Using the relation between form 
factors and mean square radius one finds
\beq
\la X^2\ra\sim-\del_{q^2} \F(q^2)|_{q^2=0}\sim \a'\log E\,.
\eeq

This agrees with the previous estimate for the square radius if we relate the 
resolution time with the energy according to $\epsilon\sim 1/E$.
The same kind of argument can be applied to the scattering of 
\emph{closed} string tachyons. In the Regge limit, the corresponding 
Virasoro-Shapiro amplitude behaves like~\cite{gsw,pol}
\beq\label{cregge}
\A_c\sim s^{2-{1\over 
2}\a'q^2}{\Gamma(\frac{1}{4}\a'q^2-1)\over\Gamma(2-{1\over 
4}\a'q^2)}\sim
{E^4\over q^2}(e^{-\med\a' q^2 \log E})^2={E^4\over q^2}\F_c^2(q^2)\,,
\eeq
giving again a mean radius that grows logarithmically with the  energy.

 The fact that the size of the string depends on the energy has important 
consequences regarding the behavior under Lorentz 
transformations~\cite{lor} 
(transverse spreading) and the black hole complementarity 
principle~\cite{comp}, but we shall not dwell on these interesting applications 
here.
For us, the moral of this simple example is that defining the size and shape 
of strings is problematic only if we assume infinite 
resolution; as long as we use finite resolution probes such as other 
strings we get well defined, finite answers.

The same method could be used, in principle, to compute form factors 
for excited states, but  there are  many different 
states at every mass level and most of then are described by very 
complicate vertex operators. As far as we know, only for states on 
the leading Regge trajectory of the open string has the method been 
used~\cite{bo}. The vertex operators for these states are quite simple, and 
the computed form factors describe  a ring distribution. This agrees with the 
classical interpretation of states on the leading Regge trajectory 
as rigidly rotating rods, with the ends describing a circumference 
of radius proportional to the energy.

Strings on the leading Regge trajectory are rather special, maximum 
angular momentum states with properties which can be very different 
from those of a typical excited string. Indeed, it has been recently 
shown that closed strings on the leading Regge trajectory have  
lifetimes proportional to their masses~\cite{ang,semi}. There are other 
families of states not on the leading Regge trajectory, but still described 
by relatively simple vertex operators, which show an even more striking 
stability, with lifetimes proportional to higher powers of the 
mass~\cite{long}. Given the  exponential growth of mass degeneracies, 
 we have plenty of ground to explore and it seems likely that 
we can still find many 
surprises as we consider other families of excited states.

On the other hand, there are different contexts,  such as the collapse of highly excited 
strings to form a black hole~\cite{hor,dam} as the mass grows above 
the correspondence point~\cite{cor}, string interactions in the 
Boltzmann equation approach to a  
Hagedorn gas~\cite{sal,sen,liz,lowe}, or the production of  `string balls' in a 
collider~\cite{rob}, where one is interested in general, statistical 
properties of typical excited strings. These general properties 
(typical sizes, decay and interaction rates, etc.) can be 
obtained by  averaging over all states in a  given mass level. This is the approach 
followed in~\cite{ayr}, where the averaged massless emission spectrum of 
highly excited strings was shown to be that of a black body at the Hagedorn 
temperature and  in~\cite{dec},  where the complete decay spectrum of 
excited strings, including branching ratios, was computed.

 In this paper we  follow the last  approach. Rather than studying 
individual excited states, we  compute the equivalent of  
`unpolarized' cross-sections, as  we  average over all the 
excited states at  given mass levels of the target string, and from 
them extract effective form factors. In a 
sense, we obtain highly detailed information about the spatial distribution of 
excited strings by directly `looking' at a statistical ensemble of 
them in a Rutherford type experiment. These form factors contain 
information not only about the geometry of excited strings, but also 
about their effective interactions with other strings.

This paper is organized as follows. In Section 2 we derive an exact 
formula for the averaged cross section corresponding to the 
scattering  of light strings by an  excited closed string and show 
that, for high energy probes,
this result can be used to define an effective form factor for the massive string. 
This form factor is explicitly evaluated in Section~3 in the case of 
very heavy target strings, and the spatial distribution of strings of 
mass $M$ is investigated in detail. Section~4 considers $1/E$ 
corrections, where $E$ is the energy of the probe, and our 
conclusions and outlook are presented in Section~5.

\section{Averaged Cross Sections and Form Factors}

We begin by  computing the averaged cross section for 
scattering of tachyons by excited \emph{closed}  strings. The reason we 
choose closed strings is twofold. On one hand, since closed strings interact 
by split-join interactions that can take place anywhere along their 
length, the cross section contains information about the spatial 
distribution of the whole string; in the case of open strings, 
 only  
  the positions of the endpoints are mapped out. On the other hand,  
the computation with open strings is  more involved, since one has to 
take into account  different orderings of the vertex operators, as well 
as group theory factors. Similarly, we choose tachyons as probes for 
the sake of simplicity. As we will see, the form factors are 
independent of the nature of the probe, which only shows up as  
polarization dependent factors in the cross-section. 

The averaged interaction rate is defined by \emph{averaging} over all initial target 
states at mass 
level $N$ and momentum $p$ and \emph{summing} over all final target states at mass 
level $N'$ and momentum $p'$. Concretely, 
\beq
{\cal R}_{c}¥(N,N',k,k')\equiv {1 \over{\cal G}_{c}¥(N)}\sum_{ \Phi_i|_{N}}\sum_{ \Phi_f|_{N'}} 
     |\A_{c}¥(\Phi_i,\Phi_f,k,k')|^2\,,
 \eeq
where the masses of the initial and final states are given by  $\a'M^2= 
4(N-1)$ and \hbox{$\a'M'^2= 4(N'-1)$} respectively,
${\cal G}_{c}¥ (N)$ is the degeneracy of the $N^\mathrm{th}$ closed 
string  mass level, 
and $k$ and $k'$ are the momenta of the incoming and outgoing 
tachyons. Since every closed string state can be written as the 
product of two open string states, it is convenient to use the well 
known holomorphic factorization property of 4-point tree 
amplitudes~\cite{gsw,pol}
\beq
{\cal A}_{c}(1,2,3,4; 4\a')=-\sin (2\pi\a' k\cdot k') {\cal 
A}_{o}(1,2,3,4;\a'){\cal A}_{o}^{*}¥(1,3,2,4;\a')\,,
\eeq
which expresses the (unique) closed string amplitude in terms of two 
inequivalent cyclic orderings for the open string vertex operators. 
Then, given that $\G_c(N)=\G_o(N)^2$, the averaged cross sections are related by
\beq\label{closed}
{\cal R}_{c}(N,N',k,k';4\a')=\sin^2 (2\pi\a' k\cdot k') {\cal 
R}_{o}(N,N',k,k';\a'){\cal R}_{o}(N,N',k',k;\a')\,.
\eeq
In what follows, we will set $\a'=1/2$, which implies that our closed string rates are valid 
for $\a'=2$. 

We can use the operator formalism to write an 
explicit formula for ${\cal R}_{o}(N,N',k,k')$,  extending the computations  carried 
out in~\cite{ayr,dec} for three-point functions. The details are presented in 
Appendix~A, where it is shown that
\beqa\label{rate}
{\cal R}_{o}¥(N,N',k,k')&=&{1 \over{\cal G}_{o}¥(N)}\int_{0}^1 dx x^{k\cdot 
p}\int_{0}^1 dy y^{k\cdot p}\oint_{C_w} 
{dw\over w} w^{-N} f(w)^{2-D} \oint _{C_v} {dv\over v}  v^ {N-N'}
 \non\\
 &\cdot&\la V_0 (-k,1) V_0 (-k',y)V_0 (k',yv) V_0 (k,xyv) \ra_{w}\,.
\eeqa
The contours $C_w$ and $C_v$ satisfy $|w|<|v|<1$ and $f(w)$ is related to the 
Dedekind $\eta$-function by
\beq\label{ded}
f(w)=\prod_{n=1}^\infty (1-w^n)=w^{-1/24} \eta\left(\frac{\ln w}{2 \pi 
i}\right)\,, 
\eeq
$D$ is the number of space-time dimensions and $V_{0}$ is the oscillator part of the 
\emph{open} string tachyon vertex operator
\beq
V(k,z)= : e^{ik.X(z)} : = V_{0}(k,z) z^{k\cdot p-1} e^{ik\cdot x}\,.
\eeq

\begin{figure}[!b]
\includegraphics[scale=.3]{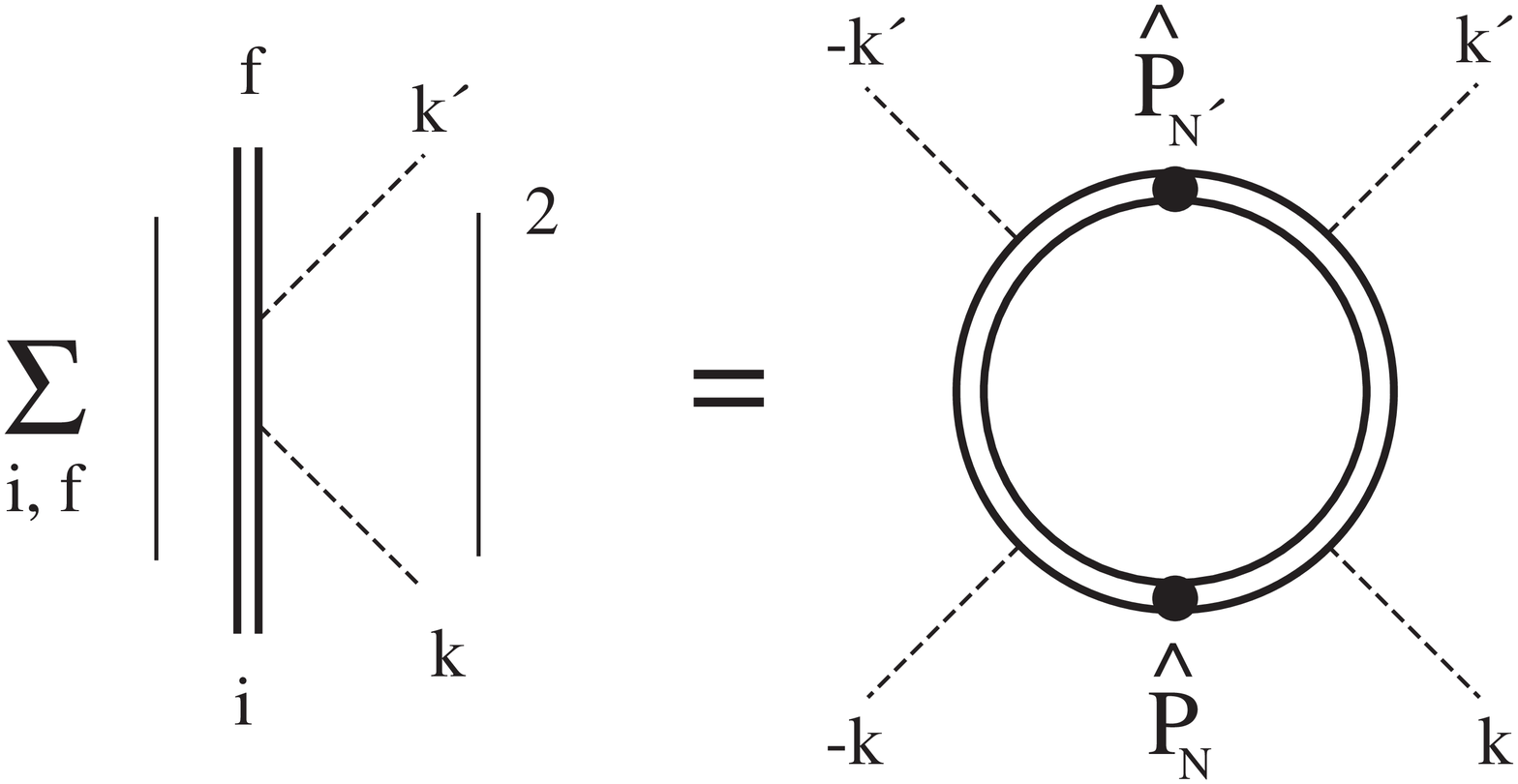}%
\caption{\label{f1} Averaged rate as projection of one-loop 
four-tachyon amplitude.}
\end{figure}

In order to evaluate~(\ref{rate}), one should 
use the oscillator part of the two point correlator on the cylinder
\beq
\la V_0 (k,z) V_0 (k',z') \ra_{w}¥=\hat\psi(z'/ z,w)^{\,k\cdot k'}
\eeq
with
\beq\label{twopoint}
\hat\psi(v,w)=(1-v)\prod_{n=1}^\infty{(1-w^nv)(1-w^n/v)\over 
(1-w^n)^2} \ .
\eeq
The $w$ and $v$ contour integrals should be done first, giving a 
linear combination of powers of $x$ and $y$. Then, the remaining $x$ 
 and $y$ integrals can be written in terms of Beta functions by 
analytic continuation.

Note that~(\ref{rate}) is closely related to the formula  giving the 
4-tachyon one-loop amplitude for the open string (see, for instance, 
Chapter~8 of~\cite{gsw}). The main differences are in the treatment of the 
zero modes (there is not loop-momentum integration, the correlator 
involves only the oscillator parts of the vertex operators) and in the 
presence of  contour 
integrals. 
Indeed, this formula can be understood as the projection of the usual 
4-tachyon loop amplitude onto (initial) states with momentum $p$ and 
level $N$ and (final) states with momentum $p'$ and 
level~$N'$, as shown in  fig.~\ref{f1}.

The computation of the differential cross-section is 
completed by using~(\ref{closed}), which gives the \emph{closed} string rate, 
and adding the appropriate phase space factors and closed string coupling 
constant $g_{c}$
\beq\label{cross}
\Bigl({d\sigma\over d\Omega}\Bigr)_{cm}={g_{c}^2\over 
2E_{p}2E|v_{p}-v|}{k^{D-3}\over (2 \pi)^{D-2}}{1\over 4E_{cm}} {\cal 
R}_{c}¥(N,N',k,k')\,, 
\eeq
where $E_{p}¥$ and $E$ are the initial energies of the target and probe 
string respectively, $v_{p}$ and $v$ are their 
velocities in the center of mass frame, and $E_{cm}=E_{p}+E$.

\subsection{Factorization and Form Factors}
Here we will show that the exact formula~(\ref{rate}) for the interaction rate 
factorizes for high energy probes at fixed momentum transfer 
$q^2=(k+k')^2$, allowing the definition  of a form factor for the probe. We begin by noting 
that 
the Regge limit~(\ref{regge}) of the
Veneziano amplitude, which one usually obtains by using Stirling formula for the Gamma
functions in
\beq
\A(k\cdot p,k\cdot k')={\Gamma(k\cdot p+1)\Gamma(k\cdot k'+1)\over\Gamma(k\cdot p+k\cdot 
k'+2)}=\int_{0}^1dx x^{k\cdot p} (1-x)^{k\cdot 
k'}\,,
\eeq
can also be obtained directly from the integral 
representation\footnote{One should note that the integral is not defined in the 
physical region where $k\cdot p<0$, and has to be analytically 
continued.}. In 
the $k\cdot p\to \infty$ limit (with  fixed $k\cdot k'$) the integral is 
dominated by the $x\sim 1$ region, and one can use Laplace  
method for integrals dominated by endpoint contributions~\cite{erde}. The change 
of variable $x=e^{-\e}$ gives
\beq\label{endpoint}
\A=\int_{0}^\infty d\e e^{-\e( k\cdot p+1)}\e^{k\cdot k'}[1+O(\e)]\sim( k\cdot 
p)^{-k\cdot k'-1} \Gamma(k\cdot k'+1)[1+O\big({1\over  k\cdot 
p}\big)]
\eeq
and generates an asymptotic expansion in $1/k\cdot p$. The first term 
of this expansion is~(\ref{regge}) (use $k\cdot k'+1=q^2/2-1$ and $-k\cdot p\sim 
s/2$). Note that this is quite different from the `hard scattering 
limit' considered by Gross collaborators~\cite{gross1,gross2} where both $k\cdot p$ 
and $k\cdot k'$ are large. In that case the integral is dominated by a 
saddle-point in the middle of moduli space, and~(\ref{regge}) can 
\emph{not} be recovered from the hard scattering approximation 
by taking the fixed $k\cdot k'$ limit, since 
the Gamma function is not reproduced. 

The same kind of argument can be used with our formula for the 
averaged rate~(\ref{rate}).
In this case, $k\cdot p=-M E$, where $M$ is target mass and $E$ is the 
probe energy in the target rest frame. For $E\gg 1$ in string units and 
fixed $q^2$, the $x$ and $y$ integrals in~(\ref{rate}) are endpoint dominated 
and one should be able to use Laplace method to generate an 
asymptotic expansion in $1/E$. To this end, make  the change of variables 
$x=e^{-\e_{1}}$, $y=e^{-\e_{2}}$, and note that the leading term in 
the expansion  corresponds to the \emph{lowest} powers of $\e_{1}$ 
and $\e_{2}$. These can be easily identified by using the following 
OPEs
\beq\label{ope}
V_0 (-k,1) V_0 (-k',y)\sim V_0 (-q,1) \e_2^{k\cdot 
k'}+\ldots\ \ ,\ \ \ 
V_0 (k',y v) V_0 (k,xyv)\sim V_0 (q,v) \e_1^{k\cdot k'}+\ldots\ \ 
\eeq
with $q=k+k'$. Upon substitution of these OPEs in~(\ref{rate}) we get
\beqa\label{approx}
{\cal R}_{o}¥(N,N',k,k')&\sim&{1 \over{\cal G}_{o}¥(N)}\int_{0}^\infty d\e_{1}¥
 e^{-\e_{1}¥ k\cdot p}
\e_{1}¥^{k\cdot k'}\int_{0}^\infty d\e_{2}¥ e^{-\e_{2}¥ k\cdot p}
\e_{2}¥^{k\cdot k'} \non\\ 
&\cdot&\oint_{C_w} {dw\over w} w^{-N} f(w)^{2-D} \oint _{C_v} {dv\over v}  v^ {N-N'}
\la V_0 (-q,1) V_0 (q,v)\ra_{w}\,. 
\eeqa
The  $1/E$ corrections to this formula come from  the 
higher powers of $\e_{1}$ and $\e_{2}$ neglected in the OPEs. The 
 $\e_{1}$ and $\e_{2}$ integrals can be  done in terms of Gamma 
 functions, giving
{\setlength\arraycolsep{1pt}{
\beqa\label{fact}
{\cal R}_{o}¥(N,N',k,k')&\sim& 
(EM)^{2-q^2}\Gamma(q^2/2-1)^2\non\\&\cdot& {1 \over{\cal G}_{o}¥(N)}\oint_{C_w} 
{dw\over w} w^{-N} f(w)^{2-D} 
\oint _{C_v} {dv\over v}  v^ {N-N'}\la V_0 (-q,1) V_0 
(q,v)\ra_{w}.  
\eeqa}
 
 This factorized form for the interaction rate suggests the following 
definition for the target effective form factor
\beq\label{fnn}
\F_{NN'}(q^2)={M^{-q^2} \over{\cal G}_{o}¥(N)} \oint_{C_w} {dw\over w} w^{-N} f(w)^{2-D} 
\oint _{C_v} {dv\over v}  v^ {N-N'}\la V_0 (-q,1) V_0 
(q,v)\ra_{w}\,.
\eeq
One  can  motivate this definition by  noting  that 
substituting~(\ref{fact}) 
with~(\ref{fnn}) into~(\ref{closed}) and
 using   the identity \hbox{$\Gamma(x)\Gamma(1-x)=\pi/\sin(\pi 
 x)$}
 gives the following expression for the \emph{closed} 
string interaction rate in the Regge limit
\beq\label{closedregge}
{\cal R}_{c}¥(N,N',k,k')\sim  
M^4\left|{\Gamma(\frac{q^2}{2}-1)\over\Gamma(2-{q^2\over 
2})}(EE')^{1-{q^2\over 2}}\right|^2|\F_{NN'}(q^2)|^2\,.
\eeq
For $N=N'$ (elastic scattering) and $M\gg E$, where target recoil can 
be neglected ($E=E'$), and for low momentum transfer $q^2\sim 0$, this gives
\beq\label{fregge}
{\cal R}_{c}¥(N,N',k,k')\sim \left|{(ME)^2\over 
q^2}\F_c(q^2)\F_{NN}(q^2)\right|^2
\eeq 
and we recognize  the square 
of the Regge 
limit of the Virasoro-Shapiro amplitude describing tachyon-tachyon 
scattering~(\ref{cregge}), with one tachyon form factor $\F_c(q^2)$ replaced by 
 $\F_{NN}(q^2)$. Thus,
  the effective form factor characterizes the difference between 
tachyon-tachyon and tachyon-probe scattering. This is analogous to 
the situation in field theory, where form factors `measure' the departure 
of real particles from ideal point-like objects. Here, what we are 
actually measuring is the departure from being `tachyon--like', or 
as point--like as a string can be.

It is important to note that, even though we have considered a very special limit (heavy 
target, elastic scattering and low momentum transfer) in order to obtain~(\ref{fregge}) and 
motivate our definition, the effective form factors are useful also 
for inelastic scattering $(N\neq N')$ and light targets. The reason is 
that~(\ref{closedregge}) holds as long as we are in the Regge limit 
of high energy probes ($E, E'\to\infty$) and fixed (not necessarily small) momentum transfer 
$q^2$, where strings are known to interact by exchanging the  
leading Regge trajectory as a whole. In this sense, the 
effective form factors describe the coupling of the target to the
leading Regge trajectory or `Reggeon'.

The formula for the effective form factor contains the correlator of 
 two \emph{off-shell} tachyon vertex operators. In fact~(\ref{fnn}) 
 is essentially the off-shell extension of the tachyon emission rate 
 for a typical open string at mass level $N$ which decays into any string 
 at mass level $N'$ (see eq.~(2.26) of~\cite{dec}). Is this related to our use 
 of tachyons as probes?
 What happens if one uses other 
 probes, such as gravitons or massive states? The only 
difference is that, instead of~(\ref{ope}), one has to compute the OPEs 
of the corresponding vertex operators, but the tachyons  still  appear as the 
leading term (lowest 
power of $\e_1$ and $\e_2$) in the r.h.s., giving rise to the 
same formula for the form factor~(\ref{fnn}). In other words, the form factors 
describe the coupling of the target string to the leading Regge 
trajectory, and this is controlled by eq.~(\ref{fnn}). Using other probes 
just changes the coupling of the probe to the the leading Regge 
trajectory. This shows up as  overall polarization dependent 
factors in~(\ref{closedregge}), but the target form factor is 
unchanged.

One can obtain the 
spatial distribution of the target by Fourier transforming the 
\emph{elastic} ($N=N'$) form factor
\beq
\rho(x)={1\over (2\pi)^{D-1}}\int d^{D-1}q\, e^{iq\cdot x} 
\F_{NN}(q^2)\,.
\eeq
A first consistency check of our definition is obtained by using
\beq\label{cor}
\la V_0 (-q,1) V_0 (q,v)\ra_{w}=\hat\psi(v,w)^{-q^2}
\eeq
and noting that
\beq
\F_{NN'}(0)=\delta_{NN'}\,.
\eeq
This implies that, for an experimenter which uses only very low 
momentum transfers \hbox{($q^2\sim 0$)}, the target will look point-like, i.e.
$\rho(x)\sim \delta (x)$.

\section{Elastic Form Factors for heavy targets}

For the rest of the paper we will study the spatial distributions of 
very heavy target strings and  be concerned only with  
\emph{elastic} form factors, which we write as $\F_{M}(q^2)\equiv
\F_{NN}(q^2)$. We  assume that $M$ is the largest scale in the problem, i.e. 
$M\gg E\gg 1$, and neglect all  corrections of order 
$O(1/M)$. In this limit the center of mass 
frame
coincides with the target rest frame where $E_{p}=M$, and the cross 
section~(\ref{cross}) can be written as 
\beq\label{crossM}
\Bigl({d\sigma\over d\Omega}\Bigr)_{cm}={g_{c}^2\over 
16 E M^2}{k^{D-3}\over (2 \pi)^{D-2}} {\cal R}_{c}¥(N,N',k,k')\,. 
\eeq

The computation of the effective form factor~(\ref{fnn}) can be
simplified
by using the well known fact that, for $N\gg 1$, the $w$-integral
\beq
\oint_{C_w} {dw\over w} w^{-N} f(w)^{2-D}\non
\eeq  
is dominated by a saddle-point at $w\approx 1$, where  the function 
$f(w)$ can be  approximated by\footnote{See~\cite{gsw} and Appendix B 
of ~\cite{dec} for 
details.}
\beq\label{fw}
\ln f(w)=\half\ln 2\pi-\half\ln\b+\frac{1}{24}\b-\frac{\pi^2}{6\b}+O
\big(e^{-4\pi^2/\b})
\eeq
with $w\equiv e^{-\beta}$.
For $w\approx 1$, $f(w)$ is a rapidly varying function, whereas the 
variation of the two-point correlator in the complete 
integral~(\ref{fnn}) is much slower. Thus the 
approximate position of the
saddle-point  can be obtained  by solving
\hbox{$N=(D-2)\partial_\b\ln f(w)$}, with the result
\beq\label{beta}
\b\approx \pi\sqrt{D-2\over 6 N}={1\over 2 M T_{H}}\,, 
\eeq
where $M=\sqrt {2N}$ and  $T_{H}$ is the Hagedorn temperature
\beq
T_{H}=\frac{1}{2\pi}\sqrt{\frac{3}{D-2}}\ .
\eeq

The solution (\ref{beta}) for $\b$ is accurate up to (relative) 
corrections of order $O(1/M)$. We will check \textit{a posteriori} that 
the shift in the saddle-point position due to the presence of the 
2-point correlator is also of order $O(1/M)$. Thus, in the saddle-point 
approximation, 
the correlator enters only   as a 
multiplicative correction.
Taking into 
account that $f(w)^{2-D}$ is the partition function for the open string
\beq
\oint_{C_w} {dw\over w} w^{-N} f(w)^{2-D}=\G_{o}¥(N)\,,
\eeq
we arrive at the following simple formula for the elastic form factor, valid 
for $M\gg 1$
\beq\label{fm}
\F_{M}(q^2)=M^{-q^2}\oint _{C_v} {dv\over v}  \la V_0 (-q,1) V_0 
(q,v)\ra_{\b}\,,
\eeq
where the correlator should be evaluated on a cylinder  with modular 
parameter $w=e^{-\b}$, with $\b$ given by~(\ref{beta}).

The correlator~(\ref{twopoint}) is written in terms 
of theta functions in Appendix~B, where the following explicit formula, 
valid for $M\gg 1$, is obtained for the elastic form factor
\beq
\F_{M}(q^2)=(2\pi T_H)^{q^2}\I(\b,q^2)
\eeq
with
\beq\label{inte}
\I(\b,q^2)\equiv {1\over \pi}\int_{0}^\pi d\xi\exp\left[q^2\left({\xi^2\over 
2\b}-\ln\cosh({\pi \xi\over\b})\right)\right]\ .
\eeq

\subsection{Evaluation of $\I(\b,q^2)$}

The integral (\ref{inte}) can not be done analytically without further 
approximations, and these depend on the range of $q^2$ we are interested.
For $q^2\gg 1/M$ ($q^2\gg \b$), the change of variable $z=\pi \xi/\b$ shows that
the quadratic term in the exponent can be neglected and we are left 
with
\beq\label{largeq}
\I(\b,q^2)\approx{\b\over \pi^2}\int_0^\infty dz\ (\cosh z)^{-q^2}=
{\b\over 2\pi^{3/2}}{\Gamma ({q^2\over 2})\over \Gamma ({q^2+1\over 
2})}\ \ \ , \ \ \ q^2\gg {1\over M}\ . 
\eeq
Evaluating the effect of the neglected  term perturbatively shows that this  formula is accurate 
up to relative corrections of order 
$O(1/Mq^2)$. One can  check the consistency of the saddle-point 
method used to evaluate the $w$-integral at the beginning of this 
section by noting that $\ln \I(\b,q^2)$ has a logarithmic dependence 
on $\b$, to be compared to the leading $1/\b$ dependence of $\ln f(w)$ 
in~(\ref{fw}).

For $q^2\simleq 1/M$, the quadratic term can not be neglected 
and~(\ref{largeq}) is no longer valid. Instead, one can use the 
approximation $\ln\cosh z\sim z-\ln 2$, valid for $z\gg 1$, giving
\beq\label{form}
\I(\b,q^2)\approx {2^{-q^2}\over\pi}\int_{0}^\pi 
d\xi\exp\left[-{q^2\over 
2\b}\,\xi(2\pi-\xi)\right]\ .
\eeq
The relative corrections to this formula are of order $O(q^4)$, and
 the approximation is valid not only for $q^2\simleq 1/M$, 
but for any $q^2\ll 1$.
This integral can be done in terms of the `imaginary error function', 
defined by
\beq
\mathrm{Erfi} (z)\equiv {2\over\sqrt\pi}\int_0^{z}dt e^{t^2}\,.
\eeq
The result is 
\beq\label{lowq}
\I(\b,q^2)\approx {\sqrt\pi\over 2z}e^{-z^2} 
\mathrm{Erfi}(z)\ \ \ ,\ \ \ z\equiv{\pi q\over 
\sqrt{2\b}}\ \ \ ,\ \ \ 0\leq q^2\ll 1\ ,
\eeq
where we have set $2^{-q^2}\approx 1$.
In this region one can set $(2\pi  T_H)^{q^2}\approx 1$ and 
\hbox{$\F_{M}(q^2)=\I(\b,q^2)$}. Now  the dependence on $\b$ 
is more involved, but one can still verify the consistency of the 
method used to evaluate the $w$-integral by noting  
that the exact equation for the saddle point is $N=(D-2)\partial_\b\ln 
f(w)-\partial_\b\ln \I(\b,q^2)$, with $\partial_\b=-(z/2\b) \partial_z$.
The contribution from $\partial_\b\ln \I(\b,q^2)$ is of order $O(1/\b)$, 
and can be neglected against  $\partial_\b\ln f(w)$, which is of order~$O(1/\b^2)$ 

It is interesting to note that~(\ref{lowq}) 
coincides with the function $g_1(q)$ describing static screening in an
electron gas at high temperature.~\footnote{I am indebted to M.~Valle for 
making me aware of this coincidence.} This is related to the real part 
of the
\textit{plasma dispersion function} $\Phi(x)$~\cite{plasma} by
\beq\label{plasma}
g_{1}(x)={2\sqrt{\pi}\over x}\;\Phi\big({x\over 
4\sqrt{\pi}}\big)\ \ \ ,\ \  \Phi(x)=2e^{-x^2}\int_0^x \d y \,e^{y^2}\,.
\eeq
The coincidence may be due to the fact that $g_{1}$ is obtained from 
a one-loop diagram evaluated in the high temperature limit where the 
electron gas behaves classically, i.e. where one can use Boltzmann distribution 
as an approximation to Fermi-Dirac (see~\cite{fet} for details). 
Our effective form factor is 
obtained as an open string one-loop  correlator~(\ref{fm}) which, due 
to the simplifications used in the large-$M$ limit, somehow seems to 
go
over to a 
(thermal) field theory one-loop correlator. On a more speculative 
level, this coincidence may suggest a correspondence between an 
ensemble of highly excited strings and a high temperature gas of 
particles or `string bits'~\cite{thorn,brane,halyo}.

Expanding~(\ref{lowq}) around $z=0$ gives
\beq\label{series}
\F_{M}(q^2)=\sum_{n=0}^\infty {(-2z^2)^n\over 
(2n+1)!!}=1-{\pi^2 \over 3\b}q^2+O(q^4)
\eeq
that, using the well known relation between the power series of the form 
factors and the distribution moments, can be used 
to obtain $\la r^{2n}\ra$ for arbitrary $n$. In particular, the mean square radius is 
given by
\beq\label{r2}
\la r^2\ra=-2(D-1)\del_{q^2} \F(q^2)|_{q^2=0}=(D-1){2\pi^2\over 
3\b}={4\pi^2\over 3}(D-1) T_H M\,.
\eeq 
As far as we know, the higher 
distribution moments have never been computed before, but the second 
moment was obtained 
long ago by Mitchell and Turok~\cite{mit} by  oscillator methods. They 
found
\beq
\la r^2\ra={D-1\over T}\sqrt{N\over 6(D-2)}\ \ \ ,\ \ \ T={1\over 
2\pi\a'}\ ,
\eeq
which agrees with (\ref{r2}) for $\a'=2$.

The ranges of validity for the two approximate expressions~(\ref{largeq}) 
and~(\ref{lowq}) overlap for $1/M\ll q^2\ll 1$. In this region, the 
integral is well approximated by
\beq\label{q2}
\I(\b,q^2)\approx {1\over 2z^2}={\b\over \pi^2  q^2}\ \ \ ,\ \ \ 1/M\ll 
q^2\ll 1\ , 
\eeq
which can be obtained  as the $q^2\to 0$ limit of~(\ref{largeq}) 
or   the 
$q^2\to\infty$ limit of~(\ref{lowq}), and has relative corrections of 
order $O(1/Mq^2)$ and $O(q^4)$.
In this regime  
 the interaction rate~(\ref{closedregge})  will be 
proportional to $M^2$, giving a cross-section~(\ref{crossM}) which is 
totally independent of the target mass! In other words, a Rutherford 
type experiment designed to explore the inner structure of highly excited 
strings would show that this structure is independent of $M$, as long 
as $1/M\ll q^2\ll 1$. We will find a very natural 
interpretation 
for this fact below.

Lastly, for $q^2\gg 1$, the $q^2\to\infty$ limit of~(\ref{largeq}) 
gives
\beq
\I(\b,q^2)\approx{\b\over \sqrt{2}\pi^{3/2}q}\ \ \ ,\ \ \  q^2\gg 1\ .
\eeq
To summarize, most of the structure of  $\I(\b,q^2)$ is concentrated 
around $q^2\simleq 1/M$. Away from this region, the integral exhibits 
a simple power-like behavior, with a cross-over from $1/q^2$ to $1/q$ 
at the string scale $q^2\sim 1$ given by~(\ref{largeq}) (see 
fig.~\ref{f2}).

\begin{figure}[!t]
\begin{center}
\includegraphics[width=7cm]{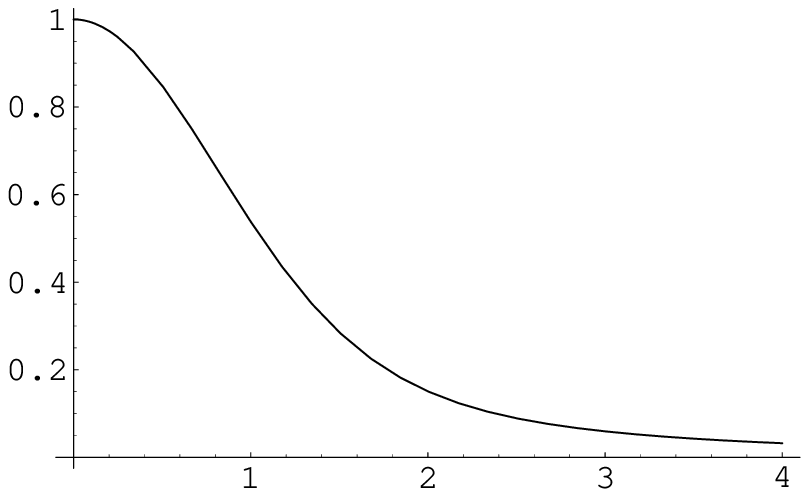}
\hspace{1cm}
\includegraphics[width=7cm]{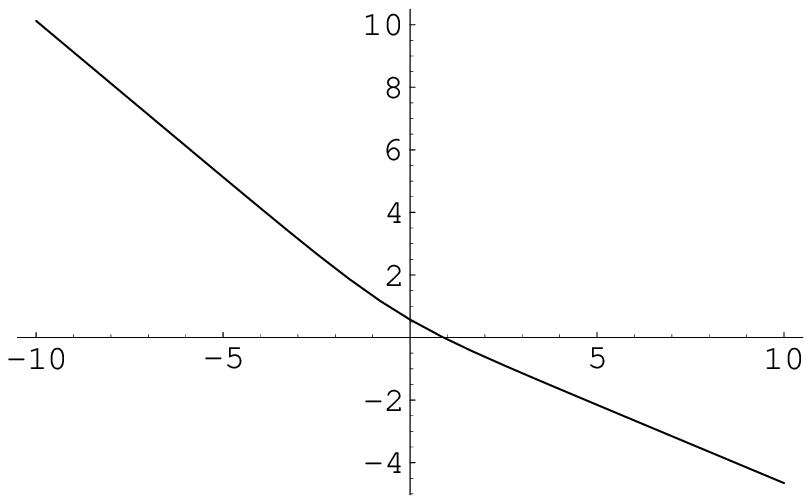}\\
\caption[fig]{\label{f2} Left: $\I(\b,q^2)$ as a function of $z$ for 
$q^2\ll 1$ as given by~(\ref{lowq}). Right: $\log\I(\b,q^2)$ \emph{versus} $\log q^2$ according 
to eq.~(\ref{largeq}). Note the change in slope from $-1$ to $-\med$ 
around $q^2=1$.}
\end{center}
\end{figure} 

\subsection{Spatial Distribution of Heavy Strings}

The spatial distribution is obtained by Fourier transforming the 
form factor. Since the radius of heavy strings is given 
by~(\ref{r2}), and most of the interesting structure arises at low~$q^2$, 
one must use~(\ref{lowq}) for the form factor. Doing the $q$-integral 
gives 
\beq\label{rho}
\rho(r)={1\over (2\pi)^{d}}\int d^{d}q\, e^{iq\cdot r} 
\F_{M}(q^2)={1\over\pi}\left({\b\over 2\pi}\right)^{{d\over 2}}\int_0^\pi 
d\xi h(\xi)^{-{d\over 2}} 
\exp\left(-{\b r^2\over 2 h(\xi)}\right)\ ,
\eeq
where $h(\xi)\equiv\xi(2\pi-\xi)$ and $d\equiv D-1$. 
Although $\rho(r)$ can not be evaluated analytically, it is obvious 
from~(\ref{rho}) that
\beq\label{norm}
\int d^{d}x\rho(r)=1\,.
\eeq
As the form factor describes the coupling of the string to the 
first Regge trajectory, and at low momentum transfer this is 
dominated by the graviton multiplet, we may assume that  $\rho(r)$ is 
proportional to the \emph{mass} distribution of the  string. The 
proportionality constant is fixed by~(\ref{norm}) and we  define 
the   mass density by   $\rho_M(r)\equiv M\rho(r)$.

In 
general,  $\rho_M(r)$ has to be computed numerically, but the behavior 
at `short distances' $1\ll r\ll \sqrt{M}$ can be obtained by setting 
$h(\xi)\approx 2\pi\xi$ in ~(\ref{rho}). The result is
\beq\label{rho2}
\rho_M(r)\approx A_{d} r^{2-d}\ \ \ ,\ \ \ A_d={\pi^{-2-d/2}\over 
8\,T_H}\,\Gamma({d\over 2}-1)
\eeq
and we see that the inner mass distribution of the string is entirely 
independent of the total mass $M$. This explains the lack of 
dependence with $M$ in the cross section that we found 
after~(\ref{q2}), which is in fact the Fourier transform of~(\ref{rho2}): The 
inner structure of the string is universal and independent of $M$, 
even at distances much larger than   
the string scale. Also note that the string has a rather `hard' core, 
as exhibited by the singular behavior of $\rho_M$ in~(\ref{rho2}).

In the opposite limit $r\gg \sqrt{M}$ the integral in~(\ref{rho}) is 
dominated by a saddle-point at $\xi=\pi$ and can be approximated by
\beq
\rho_{M}(r)\sim {A'_d\over r} \exp(-{\b r^2\over 2\pi^2})\,.
\eeq
where $A'_d=2^{-(d+1)/2}\pi^{-3(d-1)/2}\b^{(d-1)/2}$.

When $r$ is neither very large nor  small compared to the mean 
radius $R$ given by~(\ref{r2}), $\rho_M(r)$ must be obtained numerically. 
Given the singular behavior at small $r$ given by~(\ref{rho2}), it is 
convenient to consider instead the \emph{radial} density defined by 
\beq
{d m\over d r}=\rho_{M}(r)\Omega_{d-1} r^{d-1} \ \ \ ,\ \ \ 
\Omega_{d-1}={2 \pi^{d/2}\over \Gamma(d/2)}\ ,
\eeq
and the mass $m(r,M)$ within a sphere of radius $r$
\beq\label{mr}
m(r,M)=\int_o^r \d r{d m\over d r}\,.
\eeq

\begin{figure}[!t]
\begin{center}
\includegraphics[width=7cm]{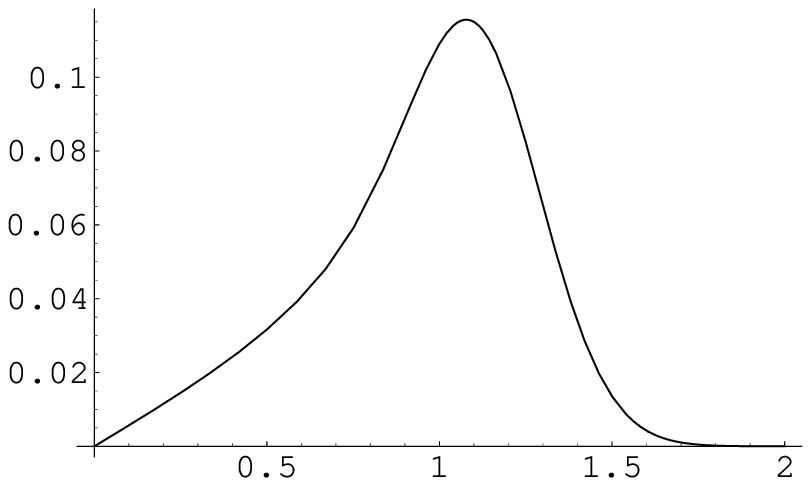}
\hspace{1cm}
\includegraphics[width=7cm]{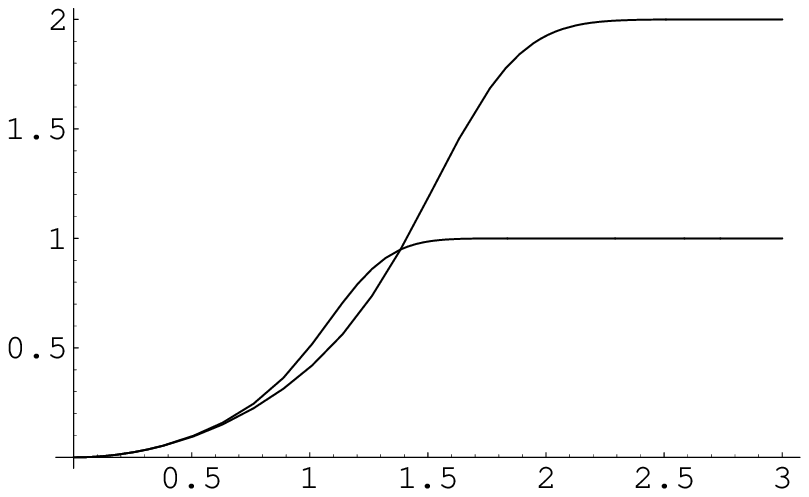}\\
\caption[fig]{\label{f3} ${1\over M}{dm\over dr}$  (left) 
and $m(r,M)/M$ (right) as  
functions of $r/R$. }
\end{center}
\end{figure}

These are represented in fig.~\ref{f3} for $d=25$ with $r$ in units of the 
mean radius $R$ and $m(r,M)$ in units of $M$. Note that the linear 
behavior for $dm/dr$ predicted  by~(\ref{rho2}) actually holds for 
$r$ up to about half the mean radius. Correspondingly, we see that the 
inner structures of strings with masses $M$ and $2M$ coincide for 
$r\simleq R/2$.


\section{Low energy corrections}

In this Section we will consider corrections to the factorized form~(\ref{closedregge}) 
for the averaged interaction rate. This was obtained by 
using~(\ref{approx}) as an approximation to the exact 
formula~(\ref{rate}). For light targets, the exponent $k\cdot p=-ME$ 
in~(\ref{rate}) is large only for $E\gg 1$, and  it is obvious that 
the  validity of the asymptotic 
approximation  requires the use of  high energy probes.

On the other hand, one might be tempted to  think that, for very heavy 
targets, the factorization 
property~(\ref{closedregge}) would hold even at low energies since,  as 
long as $EM\gg 1$, 
 the averaged 
rate~(\ref{rate}) seems to  be endpoint dominated even for low energy 
probes. In what follows\footnote{Since 
in this paper we have considered  the large mass limit only for 
\textit{elastic} form factors, for the rest of this Section we assume 
$E=E'$.} we will show that this is not the case, and 
that~(\ref{closedregge}) is the first term in an asymptotic expansion 
controlled by the small parameter $1/E$, not  $1/EM$.

The exact formula~(\ref{rate}) contains the $4$-point correlator
{\setlength\arraycolsep{0pt}
\beqa\label{exact}
\la V_0 (-k,1) V_0 (-k',y)&&V_0 (k',yv) V_0 (k,xyv) \ra_{w}=\non\\
&&\hat\psi(x)^{k\cdot k'} \hat\psi(xv)^{-k\cdot k'} 
\hat\psi(xyv)^{-2} \hat\psi(v)^{-2} \hat\psi(yv)^{-k\cdot k'} 
\hat\psi(y)^{k\cdot k'}\,.
\eeqa}
One obtains~(\ref{approx}) by making the substitutions
\beq
\hat\psi(x)^{k\cdot k'}\to \e_1^{ k\cdot k'}\ \ \ ,\ \ \ 
\hat\psi(y)^{k\cdot k'}\to \e_2^{ k\cdot k'}
\eeq
and setting $x=y=1$ in the other two-point functions. The corrections 
to~(\ref{approx}) are obtained by expanding the correlators in powers 
of $\e_1$ and $\e_2$, and this involves differentiating the two-point 
function $\hat\psi$. Each power of $\e_{1}$ or $\e_{2}$ carries a factor 
$1/EM$ from 
\beq\label{inte1}
\int_{0}^\infty d\e_{1}¥
 e^{-\e_{1}¥ k\cdot p}
\e_{1}¥^{k\cdot k'+n}=(EM)^{1-q^2/2+n}\Gamma(q^2/2-1+n)
\eeq
but, according to Appendix~B, in the large $M$ limit
\beq\label{prop}
\hat\psi(v,w)\sim\frac{\b}{\pi}\exp\left({\g^2\over 2\b}-{\g\over 2}\right)
\sin\left({\pi \g\over\b}\right)\ ,
\eeq
where $v\equiv e^{-\gamma}$. Thus, the  $n^\mathrm{th}$ derivative of $\hat\psi$
carries a factor 
of $\b^{-n}\propto M^n$, and the leading terms in the asymptotic expansion 
are powers of $1/E$. We conclude that, in general, the factorized form~(\ref{closedregge}) 
for the interaction rate is valid only for $E\gg 1$. 

This statement needs some qualifications, however,  since so far we have 
neglected the scale set by the momentum transfer $q^2$. For  $q^2\gg 1$ we have another 
large parameter, which has to be taken into account. In this limit 
one can take $k\cdot k'\sim q^2/2$ and it is obvious that the leading 
corrections will come from $\hat\psi(x)^{k\cdot k'} \hat\psi(xv)^{-k\cdot 
k'}$ and $\hat\psi(yv)^{-k\cdot k'} \hat\psi(y)^{k\cdot k'}$ 
in~(\ref{exact}). One should identify the  terms with the 
highest powers of $q^2$ per  power of $\e_1$ or $\e_2$. These are easily isolated by using
\beq
(\hat\psi(x)/\hat\psi(xv))^{q^2/2}\sim\left({\sin{\pi\e_1\over\b}\over\sin{\pi\over\b}
(\gamma+\e_1)}\right)^{q^2/2}\!\sim\; 
\left({\pi\e_1\over\b}\over\sin{\pi\gamma\over\b}
\right)^{q^2/2}\exp\bigg[-{q^2\over 
2}{\pi\e_1\over\b}\cot{\pi\gamma\over\b}\bigg]\ . 
\eeq
where we have exponentiated $(1+x)^a\sim e^{xa}$. Comparing this to~(\ref{inte1}), it is obvious that the resummed 
corrections in the $q^2\gg 1$ limit are equivalent to the 
substitution
\beq
k\cdot p\to k\cdot p\left(1+\med {q^2\over k\cdot p}{\pi\over\b}\cot 
{\pi\gamma\over\b}\right) 
\eeq 
which gives an overall multiplicative correction to the $\e_1$-integral
\beq
(EM)^{-q^2/2}\to(EM)^{-q^2/2}\exp\left(-{1\over 4}{q^4\over k\cdot p}{\pi\over\b}\cot 
{\pi\gamma\over\b}\right)\ .
\eeq
Taking into account an identical contribution from the 
$\e_2$-integral and making the change of variables $\g=\b/2+i\xi$ as in Appendix~B, 
we see that instead of~(\ref{inte})  we now have
\beq
\I(\b,q^2)\to {1\over \pi}\int_{0}^\pi d\xi\exp\left[q^2\left({\xi^2\over 
2\b}-\ln\cosh({\pi \xi\over\b})\right)+{i\over 2}{q^4\over k\cdot 
p}{\pi\over\b}\tanh 
{\pi\xi\over\b}\right]\ .
\eeq
In the $q^2\gg 1$ limit this integral can be approximated by a 
gaussian, and it is easy to see that the effect of the new term in 
the exponent is an overall factor
\beq
\I(\b,q^2)\to \I(\b,q^2)\exp\left(-{ \pi^2q^6\over 8 (k\cdot 
p)^2\b^2}\right)\sim \I(\b,q^2)\exp\left(\!-{ \pi^2 T_H^2q^6\over 
2E^2}\right)\ ,
\eeq 
where we have used $k\cdot p=-EM$ and the value of $\b$ given 
by~(\ref{beta}).
Thus, for large momentum transfer it is not enough to  have $E\gg 
1$, and  
the condition on the energy becomes $E\gg q^3$. This implies very 
small scattering angles and is typical of the Regge region, 
modified in this case by the presence of a large mass.

\section{Discussion}

In this paper we have used factorization and Regge limit 
asymptotics to obtain effective form factors for typical excited \emph{closed} 
strings. These are written  
in terms of two point functions for \emph{open} string tachyons 
on a cylinder~(\ref{fnn}) and, for very 
heavy targets and elastic scattering, take the simple form
\beq\label{fm5}
\F_{M}(q^2)=M^{-q^2}\oint _{C_v} {dv\over v}  \la V_0 (-q,1) V_0 
(q,v)\ra_{\b}\,.
\eeq
We have done this by following  an old suggestion to define physically sensible string 
form factors by extracting them from scattering 
amplitudes~\cite{size,bo,comp}. 
This way  the problem of  infinite size is overcome 
by using light strings as finite resolution probes, which effectively 
cut-off  the infinite sums over  string modes. Moreover, the 
factorization process gives  a well defined prescription for the  
 off-shell correlator in~(\ref{fm5}). In this regard, we must  remember that~(\ref{fm5}) is
\emph{not} conformally invariant, and is valid only for a particular 
parametrization of the world-sheet, namely the one where
 the two point function is given by~(\ref{twopoint}).

The elastic form factor $\F_M(q^2)$ has a rather 
simple behavior and for $q^2\ll 1$ can be written in closed form 
in terms of the `plasma dispersion function' (eqs.~(\ref{lowq}) and 
~(\ref{plasma}))  Since this is 
associated with static charge screening in a hot plasma~\cite{fet}, its appearance 
in this context is somewhat surprising, and may be related to the 
possibility of a  
`string bit' interpretation for the statistical  ensemble of 
excited strings~\cite{thorn,brane,halyo}.

Fourier transforming the formula for $\F_M(q^2)$ valid for $q^2\ll 1$ 
gives the mass distribution for $r\gg 1$, i.e. at low resolution 
with respect to  the string scale. This is shown in fig.~3, which can 
be considered as a `picture' of the string ensemble obtained by using 
light string probes in a Rutherford type experiment. The fact that 
the interior of the string is independent of the mass indicates that, 
as the mass is increased, the string grows by piling up new shells  
on top of another, while the core is unchanged.

The form of the density $\rho(r)$ given by~(\ref{rho}) 
implies the following scaling property for the mass $m(r,M)$ within a 
sphere of radius defined by~(\ref{mr})
\beq\label{scale}
m(r,M)=M\, m({r/ \sqrt{M}},1)\ .
\eeq
This kind of scaling is characteristic of random walks, and in 
particular implies \hbox{$\la r^2\ra\propto M$}. This property of the 
second moment  is usually taken to mean 
that, in some sense, highly excited strings are random 
walks~\cite{sal,mit}. The highly detailed information about the spatial distribution 
of excited strings  obtained in 
this paper could be 
used to test this idea.
In particular, one could compare  the higher moments $\la r^{2n}\ra$, which are not uniquely 
determined by scaling but can be obtained from the Taylor 
expansion~(\ref{series})
of $\F_M(q^2)$, to the ones predicted by some concrete realization of 
the random walk model. 

Although we have only considered elastic scattering by heavy targets 
in this paper, the analysis could be extended to inelastic processes, 
giving information about  the dynamical response of highly excited  strings.
One possible application is the study of thermalization in a  Hagedorn 
gas. In particular, one could investigate how the energy of very energetic 
light strings gets redistributed among internal  (string excitations) and 
external (kinetic energy) degrees 
of freedom after  colliding with slow heavy strings.

In this paper we have studied the bosonic string for the sake of 
simplicity, but the technique could be extended  to the 
superstring, although we do not expect qualitatively new results. 
Another question that we have omitted entirely is the effect of 
higher loop corrections. Since we were primarily interested in the 
internal structure of \emph{free} excited strings, we have implicitly 
assumed that the coupling constant is so small that the tree 
level results  can be trusted. Going beyond this approximation 
 is a  hard problem, but one might 
try something along the lines of~\cite{muz,eik}. According to our 
analysis in Section~4, factorization holds in the Regge limit where 
one could try to do an eikonal resummation of leading loop effects, 
although this may be hampered by the fact that  effective form 
factors, unlike ordinary amplitudes,  contain no phase information.

\begin{acknowledgments}
It is a pleasure to thank  I.L.~Egusquiza, R.~Emparan, M.~Uriarte, 
  M.A.~Valle-Basagoiti and M.A.~V\'azquez-Mozo for useful and
interesting discussions. This work  has been  supported in part by
the Spanish Science Ministry under Grant FPA2002-02037 and by  University of 
the Basque Country Grant UPV00172.310-14497/2002. 
\end{acknowledgments}

\appendix

\section{}
Here we compute the averaged open string rate. This can be written as 
\beqa
{\cal R}_{o}(N,N',k,k')&\equiv& {1 \over{\cal G}_{o}¥(N)}\sum_{ \Phi_i|_{N}}\sum_{ \Phi_f|_{N'}} \big|
 \langle \Phi_f |  V(k')\D V(k)|\Phi_i \rangle \big|^2\non\\
 &=&{1 \over{\cal G}_{o}¥(N)}\sum_{ \Phi_i|_{N}}\sum_{ \Phi_f|_{N'}}\langle \Phi_i |  V\da (k)\D 
 V\da(k')|\Phi_f \rangle \langle \Phi_f |  V(k')\D V(k)|\Phi_i 
 \rangle\,,
\eeqa 
where $V$ is the tachyon vertex operator and $\D$ is the open string propagator
\beq\label{}
\D= \int _0^1 {dx\over x} x^ {L_{0}-1}\ \ \ , \ \ \ L_{0}= \hat 
N+{p^2 \over 2}\ .
\eeq
Then the sums are converted into a trace with the help of the level 
projection operators
\beq\label{cincuentaiuna}
\hat P_N= \oint _C {dz\over z} z^ {\hat N-N}\ \ \ , \ \ \ \hat 
N=\sum_{n=1}^\infty \a_{-n}\cdot\a_{n}\ .
\eeq
The result is
\beq\label{dos}
{\cal R}_{o}¥(N,N',k,k')={1 \over{\cal G}_{o}¥(N)} \oint_C {dz\over z} z^{-N  }  \oint _{C'} {dz'\over z'} 
z'^{-N ' }{\rm Tr}\ \big[z^{ \hat N } V\da(k)\D V\da(k')  {z'}^{ \hat N } 
V(k')\D V(k)\big]
\eeq
where $C$ and $C'$ are small contours around the origin. Using
\beq
z^{ L_{0} }\, V (k,1) {z}^{ -L_{0}}=V(k,z)\ ,\ \ \ \ \ \ z^{\hat N }\, 
V_{0}¥ (k,1) 
{z}^{ -\hat N}=V_{0}(k,z)\   ,
\eeq 
where $V_{0}$ is the oscillator part of the tachyon vertex operator
\beq
V(k,z)=: e^{ik.X(z)}:=V_{0}(k,z) z^{k\cdot p-1} e^{ik\cdot x}
\eeq
and making the change of variables  $v=z'$ and $w=xyzz'$ gives
\beqa
{\cal R}_{o}¥(N,N',k,k')&=&{1 \over{\cal G}_{o}¥(N)}
\int_{0}^1 dx x^{k\cdot p}\int_{0}^1 dy y^{k\cdot p} \oint_{C_w} 
{dw\over w} w^{-N}  \oint _{C_v} {dv\over v}  v^ {N-N'}
\non\\
&\cdot&{\rm Tr}\ \big[
V_0 (-k,1) V_0 (-k',y) V_0 (k',yv) V_0 (k,xyv) w^{ \hat N }\big]\,,
\eeqa
where $p$ is the initial momentum of the target string. This can be 
converted into a 4-point correlator on a cylinder of modular 
parameter $w$ by using the 
identity\footnote{In order to obtain this 
result one should also introduce  ghosts  and 
add the corresponding contribution to the number operator in the 
projectors~(\ref{cincuentaiuna}).}
{\setlength\arraycolsep{-2pt}
\beqa
{\rm Tr}\ \big[
V_0 (-k,1)&& V_0 (-k',y)V_0 (k',yv)V_0 (k,xyv) w^{ \hat N }\big]\non\\ 
&=&\;\;f(w)^{2-D} \la V_0 (-k,1) V_0 (-k',y)V_0 (k',yv) V_0 (k,xyv) 
\ra_{w}\ ,
\eeqa}
where $f(w)$ is given by~(\ref{ded})
and $D$ is the number of space-time dimensions. 

The final formula for 
the averaged rate is
{\setlength\arraycolsep{0pt}
\beqa
{\cal R}_{o}¥(N,N',k,k')={1 \over{\cal G}_{o}¥(N)}\int_{0}^1 dx x^{k\cdot 
p}&&\int_{0}^1 dy y^{k\cdot p}\oint_{C_w} 
{dw\over w} w^{-N} f(w)^{2-D} \oint _{C_v} {dv\over v}  v^ {N-N'}
 \non\\
 &\cdot&\la V_0 (-k,1) V_0 (-k',y)V_0 (k',yv) V_0 (k,xyv) \ra_{w}.
\eeqa}

\section{}
In order to evaluate the integral in~(\ref{fm}) one should use (\ref{cor}) 
together with
\beq\label{rel}
\hat\psi(v,w)=\sqrt{v}\exp\left( {-\ln^2 v\over2 \ln w}\right)\ 
\psi(v,w)\,,
\eeq
where $\psi(v,w)$ is the scalar correlator on the cylinder
\beq\label{cuarentaicinco}
\la \Xm (1) \Xn (v) \ra=-\eta^{\mu\nu}\ln\psi(v,w)
\eeq
which  is related to the Jacobi $\vartheta_{1}$ function 
by \cite{gsw}
\beq\label{auno}
\psi(v,w)=\frac{2\pi 
i}{\t}\,\frac{\vartheta_{1}(\nu|\t)}{\vartheta'_{1}(0|\t)}\ \ \ ,\ \ 
\t\equiv-\frac{2\pi i}{\ln w},\ \ \ ,\ \ \nu\equiv\frac{\ln v}{\ln w}\ . 
\eeq
Using the product formula  for the theta function  yields the 
following  expression  
\beq
\psi(v,w)=-\frac{2\pi}{\ln q}\sin \pi \nu\prod_{1}^\infty\frac{1-2q^{2n}\cos 
2\pi\nu+q^{4n}}{(1-q^{2n})^2}\ \ \ \ ,\ \ \ 
q=e^{i\pi\t}=e^{-2\pi^2/ \b}\  .
\eeq
The infinite 
product in this formula is equal to one, up to corrections which are exponentially 
suppressed in the large $M$ limit. Using (\ref{rel}) and defining
$v\equiv e^{-\g}$ gives
\beq
\hat\psi(v,w)\sim\frac{\b}{\pi}\exp\left({\g^2\over 2\b}-{\g\over 2}\right)
\sin\left({\pi \g\over\b}\right)\ .
\eeq

We are now  ready to give an explicit formula for the $v$-integral. Since $C_v$ 
can be any  circle with $1>|v|>|w|$, in terms of the 
variable $\g$ the integral runs along $\g=c+i\xi$, where $0<c<\b$  and $\xi$ is a real variable. 
Using the convenient choice $c=\b/2$ gives
\beq
\oint _{C_v} {dv\over v}  \la V_0 (-q,1) V_0 
(q,v)\ra_{\b}={1\over 
2\pi}\left({\b\over\pi}\right)^{-q^2}e^{\b q^2/8}\int_{-\pi}^\pi d\xi\exp\left[q^2\left({\xi^2\over 
2\b}-\ln\cosh({\pi \xi\over\b})\right)\right]\ .
\eeq
As we are assuming that the momentum transfer $q^2$ is very small 
compared to the mass $M$ of the target, we can set $\b q^2/8\approx 
0$, and using~(\ref{beta}) one has the following simple expression 
for the
 elastic form factor 
\beq
\F_{M}(q^2)=(2\pi T_H)^{q^2}\I(\b,q^2)
\eeq
where
\beq
\I(\b,q^2)\equiv {1\over \pi}\int_{0}^\pi d\xi\exp\left[q^2\left({\xi^2\over 
2\b}-\ln\cosh({\pi \xi\over\b})\right)\right]\ .
\eeq



\begin{thebibliography}{99}

\bibitem{had} L.~Susskind, \textit{Structure of Hadrons implied by 
Duality}, \prd{1}{1970}{1182}.

\bibitem{size} M.~Karliner, I.~Klebanov and L.~Susskind, \textit{Size 
and Shape of Strings}, Int. J. Mod. 
Phys. {\bf A3} (1988) 1981.

\bibitem{gsw} M.~Green, J.~Schwarz and E.~Witten, \emph{Superstring Theory}, 
Vols. I and II, Cambridge~1987.

\bibitem{pol} J.~Polchinski, \emph{String Theory}, Vols. I and II, 
Cambridge~1998.

\bibitem{comp} L.~Susskind, \textit{String Theory and the Principle 
of Black Hole Complementarity}, \prl{71}{1993}{2367}.

\bibitem{lor} L.~Susskind, \textit{Strings, Black Holes and Lorentz 
Contraction}, \prd{49}{1994}{6606}.

\bibitem{bo} D.~Mitchell and B.~Sundborg, \textit{Measuring the Size 
and Shape of Strings}, Nucl. Phys. {\bf B349} 
(1991) 159.

\bibitem{ang} R.~Iengo and J.~G.~Russo, \textit{The Decay of Massive 
Closed Superstrings with Maximum Angular Momentum}, 
\jhep{11}{2002}{045} [hep-th/02102245].

\bibitem{semi} R.~Iengo and J.~G.~Russo, \textit{Semiclassical Decay 
of Strings with Maximum Angular Momentum}, \jhep{03}{2003}É{030} 
[hep-th/0301109].

\bibitem{long}  R.~Iengo and J.~G.~Russo, \textit{Decay of Long-Lived 
Closed Superstring States} [hep-th/0310283].
 
\bibitem{hor} G.~T.~Horowitz  and J.~Polchinski, \textit{Self 
Gravitating Fundamental Strings}, \prd{57}{1998}{2557}, 
[hep-th/9707170].

\bibitem{dam} T.~Damour and G.~Veneziano, \textit{Self-gravitating 
Fundamental Strings and Black Holes}, Nucl. Phys. {\bf B568} 
(2000) 93 [hep-th/9907030].

\bibitem{cor} G.T.~Horowitz and J.~Polchinsky, \textit{A 
Correspondence Principle for Black Holes and Strings}, Phys. Rev. {\bf D55} 
(1997) 6189 [hep-th/9612146].

\bibitem{sal} P.~Salomonson and B.~Skagerstam, \textit{On Superdense 
Superstring Gases: A Heretic String Model Approach}, Nucl. Phys. {\bf B268} 
(1986) 349.

\bibitem{sen} I.~Senda, \textit{A Nucleation Model of Hadrons Based 
on the Dual String}, \plb{263}{1991}{270}.

\bibitem{liz} F.Lizzi and I.~Senda, \textit{The Nucleation Model of 
the Hagedorn Phase Transition}, \npb{359}{1991}{441}.

\bibitem{lowe} D.A.~Lowe and L.~Thorlacius, \textit{Hot String Soup: Thermodynamics 
of Strings near the Hagedorn transition}, Phys. Rev, {\bf D51} (1995) 
665 [hep-th/9408134].

\bibitem{rob} S.~Dimopoulos and R.~Emparan, \plb{256}{2002}{393} [hep-ph/0108060].

\bibitem{ayr} D.~Amati and J.G.~Russo, \textit{Fundamental Strings as 
Black Bodies},  Phys. Lett. {\bf B454} (1999) 
207 [hep-th/9901092].

\bibitem{dec} J.~L.~Ma\~nes, \textit{Emission Spectrum of Fundamental 
Strings: An  Algebraic Approach}, \npb{621}{2002}{37} [hep-th/0109196].

\bibitem{erde} A.~Erdelyi, \textit{Asymptotic Expansions}, Dover 1987.

\bibitem{gross1} D.~J.~Gross and P.~F.~Mende, \textit{String Theory 
beyond the Plank Scale}, \npb{303}{1988}{407}.

\bibitem{gross2} D.~J.~Gross and J.~L.~Ma\~nes, \textit{High Energy 
Behavior of Open String Scattering}, \npb{326}{1989}{73}.

\bibitem{plasma} D.~B.~Fried and S.~D.~Conte, \textit{The Plasma 
Dispersion Function}, Academic Press, New York~1961.

\bibitem{fet} A.~L.~Fetter and J.~D.~Waleka, \textit{Quantum Theory of 
Many-Particle Systems}, McGraw-Hill~1971.

\bibitem{thorn} O.~Bergman and C.~B.~Thorn, \textit{String Bit Models 
for Superstring}, \prd{52}{1995}{5980} [hep-th/9506125].

\bibitem{brane} E.~Halyo, A.~Rajaraman and L.~Susskind, 
\textit{Braneless Black Holes}, \plb{392}{1997}{319} [hep-th/9605112].

\bibitem{halyo} E.~Halyo, \textit{Gravitational Entropy and String 
Bits on Stretched Horizons} [hep-th/0308166].

\bibitem{mit} D.~Mitchell and N.~Turok, \textit{Statistical 
Properties of Cosmic Strings}, Nucl. Phys. {\bf B294} (1987) 
1138.

\bibitem{muz} I.~J.~Muzinich and  M.~Soldate, \textit{High-energy 
Unitarity of Gravitation and Strings}, \prd{37}{1988}{359}.

\bibitem{eik} D.~Amati, M.~Ciafaloni  and G.~Veneziano, 
\textit{Classical and Quantum Gravity Effects from Planckian Energy 
Superstring Collisions}, \ijmpa{3}{1988}{1615}.




\end{thebibliography}
\end{document}